\documentstyle[12pt,BIB]{report}    % Specifies the document style.
\renewcommand{\theequation}{\thesection.\arabic{equation}}
\newcommand{\be}{\begin{equation}}
\newcommand{\ee}{\end{equation}}
\def\lsim{\raise0.3ex\hbox{$<$\kern-0.75em\raise-1.1ex\hbox{$\sim$}}}
\def\gsim{\raise0.3ex\hbox{$>$\kern-0.75em\raise-1.1ex\hbox{$\sim$}}}

\title{ The Production of the Exotic Atoms $\pi^+\pi^-$, $K^+\pi^-$
 and $K^+K^-$.}
\author{S. Wycech\thanks{Bitnet address:"WYCECH@FUW.EDU.PL"},\\
Soltan Institute for Nuclear Studies,
Warsaw, Poland \\ and \\
A.M. Green\thanks{Bitnet address:"GREEN@FINUHCB"},\\
 Research Institute for Theoretical Physics,  University
of Helsinki, Finland}
\begin{document}

%\large
%\pagestyle{empty}

\maketitle

\begin{abstract}
Estimates are made of the signals to be expected in the production of the
exotic atoms Pionium ($\pi^+\pi^-$), Kaonium ($K^+K^-$) and also
$K^+\pi^-$ in pp, pd and $e^+e^-$ reactions. Such experiments are now being
undertaken or contemplated at  CELSIUS, COSY, Indiana and SATURNE.

\vskip 0.5cm

\end{abstract}
\newpage

\section{ Introduction}
\setcounter{equation}{0}
Two-meson systems ($M_1M_2$) are of great theoretical interest since they
are the simplest involving the interaction between two hadrons. For example,
at a microscopic level, $M_1M_2$ is thought to be well described in terms
 of quarks as $(q\bar{q})(q\bar{q})$ -- the next degree of quark complication
after that of a single meson ($q\bar{q}$) or a baryon ($qqq$). Also at low
energies some two-meson systems exhibit special symmetries such as
chiral symmetry in the case of two pions. In spite of this theoretical
interest, the amount of available experimental data is limited to the
following.

a) In the $\pi\pi$ system, phase shifts are known upto an
energy of $\approx$1.5GeV in the centre-of-mass -- extracted
from the final state interaction in, for example, reactions such
as\cite{Martin}
\[ \pi N\rightarrow (\pi\pi)N. \]

In addition, the low energy parameters (i.e. scattering lengths and effective
ranges) can also be studied by the $K_{e4}$ decays e.g.
$K^-\rightarrow \pi^+\pi^-e\bar{\nu}$. The standard compilation is that
of ref.\cite{Dumb} with a more recent review being ref.\cite{Ochs}.

b) In the $K\pi$ system, phase shifts are known upto $\approx 1.5$GeV
\cite{Aston}. For a recent review of the theoretical and experimental status
see ref.\cite{Lohse}.

c)  In the $K\bar{K}$ system very little is known experimentally, the main
source of
information being the indirect process $\pi\pi \rightarrow K\bar{K}$ -- see
refs.\cite{Lohse,fdecay,Can} for reviews of the current theoretical and
experimental status.

So far the study of two-meson {\em atomic} states has not been
feasible experimentally. However, there is a hint that they may be formed
in high-energy proton-nucleus collisions \cite{afan}.
In principle, the energy levels and, in particular,
the lifetimes of such systems,
even though dominated by the Coulomb interaction, could give valuable
information about the much shorter range strong interaction.
For anti-protonic atoms ($\bar{p}A$), this field has, for a long time, been
the object of much research both theoretically and experimentally
\cite{th,exp},
and has successfully produced constraints on the basic $\bar{p}-$nucleon
interaction.

Recently, proposals has been made \cite{Nann,Bo} to produce pionium
$(\pi^+\pi^-)$ in the reaction
\be
\label{1.1}
 p+d \rightarrow \  ^3He+(\pi^+\pi^-)_{Atom}.
\ee
It is mainly this experiment that will be discussed in some detail in this
 article. However, the model is easily extended to other similar reactions
-- for example -- the process $pp\rightarrow pp\pi^+\pi^-$(at threshold)
recently proposed at the COSY-Julich synchroton ring \cite{Oelert}.
In addition, the formalism developed for these reactions
is also applicable to other exotic atoms such as $K^+K^-$
 or $K^+\pi^-$. These atoms
may also be accessible in the near future -- for example, there are indications
of a signal at the $K\bar{K}$ threshold in the recent
$pd \rightarrow \ ^ 3HeX$ at Saturne \cite{Siebert}. In section 2 some
basic features of exotic atoms are given.
In section 3, the coalescence model is introduced to estimate the relative rate
of atom $(M^+M^-$) production compared with the uncorrelated pair production
$M_1M_2$. For the reaction in eq.(\ref{1.1}), this is basically the ratio of
the phase space factors $L^2/L^3$, where $L^2$ is the two-body phase
space $L^2[ \ ^3He,(\pi^+\pi^-)_{atom}]$ and
$L^3$ is the three body-phase space $L^3[ \ ^3He,\pi\pi]$.
Each of these is also weighted by an appropriate production amplitude for
the $(\pi^+\pi^-)_{atom}$ and
uncorrelated $\pi\pi$. These two amplitudes are derived in section 4 and the
atomic signal in the cross-section for the recoil $^3$He energy distribution
is calculated for that kinematics which probes the energy of the expected
atom. In section 5 the formation of pionium and kaonium
 in  2p-states is discussed and some conclusions are given in section 6.
\vskip 0.5cm

\section{ Basic features of Strange Atoms}
\setcounter{equation}{0}

Some of the basic properties of $(\pi^+\pi^-)$, $(K^-\pi^+)$ and
$(K^-K^+)$ atoms are summarised in table 1. Pionium has been discussed
earlier in a number of papers\cite{pionium} and experimental searches
have either been attempted\cite{afan} or proposed\cite{Nann}. In
this paper some additional properties of the simplest strange atoms are
also calculated.
Table 1 collects together the Bohr radii $(1/\mu \alpha)$ and some binding
energies $[\mu\alpha^2/2(n+l+1)^2]$ of these atoms, where $\mu$ is the
appropriate reduced mass. However, the quantity of special interest is the
lifetime, since it is connected to meson-meson scattering lengths by the
relation given in eq.(\ref{A7}) of appendix 1 as
\be
\label{2.1}
\Gamma=-4\mu^2\alpha^3 {\rm Im} A_{cc}.
\ee
In the $\pi^+\pi^-$ and $K^+\pi^-$ cases, Im$A_{cc}$ is generated by
isospin symmetry breaking through the meson mass differences within
the $\pi$ and $K$ multiplets.  It may be expressed in terms of the atomic
decay momentum and the combinations of the isospin scattering lengths given
in table A1 and eq.(\ref{A2}) in appendix 1. The lifetimes given in table 1
 are calculated using

$a_0=-0.26 \ \ a_2=0.02 \ \ \pi\pi-{\rm case}$

$a_1=-0.22 \ \ a_3=0.06 \ \ K\pi-{\rm case}$

\noindent The units are $1/m_{\pi}$ and the signs of the $a$'s are {\em
opposite}
to those used in ref.\cite{Dumb} -- the source of these numbers. At this stage
there is no direct data on the $K^+K^-$ scattering lengths. This system is
half isospin 0 and half isospin 1. The $I=0$ component is strongly coupled
 to the $f_0(975)$ meson and decays into pions\cite{fdecay}. Simple
estimates based on the assumption that $f_0$ is a $K\bar{K}$ quasibound state
yield Im$A_0\approx-1$ fm and similar estimates follow from
phenomenological analyses\cite{Can}. The $A_1$ length is not known, but
it could well be equally large due to $\pi\eta$ coupling via the $a_0(983)$
resonance. For illustration, the width in table 1 is, therefore, calculated
using $K_{oo}=0.328$, $K_{cc}=2.53$ and  $K_{co}=0.948$ all in fm. --
 reasonable values  suggested in ref.\cite{Can}. These are obtained
following ref.\cite{Can} by fixing the $\pi - \pi$ resonance at the
$f_0$ energy (which yields $K_{oo}$) and, from eq.(\ref{A2}),
result in $A_{cc}=1.48-i1.33$fm.

Lifetimes of the atomic 1s-states are too short to be measured directly by
electronic
methods. Another way to possibly measure such lifetimes is to observe mesonic
decay modes and compare it with $\gamma\gamma$ decays, which are exactly
calculable. The main problem with this method is the large background
 due to the direct production of the decay mesons. This problem is discussed
in the next section.

Another method would be to create relativistic atoms or atoms in 2p-states.
This latter possibility is discussed in section 5.

\section{ The coalescence model}
\setcounter{equation}{0}

The production of atoms in the coalescence model is assumed to be given
essentially
 by the atomic wave function $\psi$. Let the $M^+M^-$ meson-pair
production amplitude be $F({\bf P},{\bf q},{\bf p_i},{\bf p_f})$,
where ${\bf P}$ and ${\bf q}$ are the total and relative pair momenta,
while ${\bf p_i}$ and ${\bf p_f}$ are the momenta of other particles in the
initial and final  states. The amplitude to produce atoms in a state
$\psi$ is then
\be
\label{An}
F({\bf P},{\bf p_i},{\bf p_f})=
\int F({\bf P},{\bf q},{\bf p_i},{\bf p_f})
\frac{\tilde{\psi}({\bf q})}{(2\pi)^{3/2}}d{\bf q}.
\ee
In general, the momenta involved in $\psi({\bf q})$ are small on the usual
short range scale of meson production mechanisms. Hence for s-states
\be
\label{Ans}
F_{s}=F({\bf P},{\bf q}=0,{\bf p_i},{\bf p_f})
\int \frac{\tilde{\psi}({\bf q})}{(2\pi)^{3/2}}d{\bf q}=
F({\bf P},0,{\bf p_i},{\bf p_f})
\psi({\bf r}=0),
\ee
and so the amplitude is related to the atomic wave function at zero
range i.e. with the normalisation constant. The relative rate of atom
production/pair production is then given by phase space as
\be
\label{3.3}
\frac{W(atom)}{W(pair)}=
\frac{\int |F({\bf q}=0)|^2dL^{n+1}|\psi(0)|^2/(2\mu_{M^+M^-})}
{\int |F({\bf q})|^2dL^{n+2}},
\ee
where, for example, $L^{n+1}$ denotes the combined phase space for the atom
and the other $(n)$ final state particles. The reduced mass factor
$\mu_{M^+M^-}$ is a remnant of the relativistic phase space element, which
for a single particle is $d{\bf p}/[2E(2\pi)^3]$. However, as shown in
appendix 2, eq.(\ref{3.3}) has its limitations, since the atom is unstable.

Using the above expressions for the reaction in eq.(\ref{1.1}), the
coalescence model yields relative pionium
production rates of $2 \cdot 10^{-3}$ at $E-E_{threshold}\approx 1$MeV dropping
to $8 \cdot 10^{-5}$ at 10MeV. Relative rates for $K\bar{K}$ atom production
are higher, mainly because of the $(m_K/m_{\pi})^2\approx 10$ factor that
enters from the $|\psi(0)|^2/\mu$ term in eq.(\ref{3.3}). The rates are
small due to the large atomic radii and are clearly unmeasurable.
 However, some improvement may be obtained
with subthreshold experiments -- as described in the next section.
\vskip 0.5cm

\section{ Subthreshold production}
\setcounter{equation}{0}
In this section is discussed the atomic signal expected to occur
on the $^3$He recoil energy distribution in the $p+d$ reaction of
eq.(\ref{1.1}).
 The notation is adapted to the
pionium case, but the formalism is applicable to kaonium and to other
cases including heavier targets. To be more specific
an experiment is discussed for creating pionium below the
$\pi^+\pi^-$ production threshold \cite{Nann} by means of the  process

\vskip 0.5 cm

\begin{tabular}{ccccc}
\label{reactions}
 p+d &$\rightarrow$ & $^3He+(\pi^+\pi^-)_{1s}$& $\rightarrow $
& $^3He+\pi^0\pi^0$ \\
     &            &                         & $\rightarrow$ &
$^3He+\gamma\gamma$
\end{tabular}

\vskip 0.5 cm

The pionium decays predominantly into $\pi^0\pi^0$ and these decaying pions
would be observed on the background of neutral pions from the direct
$\pi^0\pi^0$ production. The aim of this section is to calculate the
atomic signal in the $\pi^0\pi^0$ channel as reflected in the $^3$He recoil.
In a model calculation this
signal, relative to the overall rate, is evaluated.

The basic atom formation amplitudes are given in figs. 1a) and 1b). These
will interfere with the large background of $\pi^0\pi^0$ production from
the direct process of fig.1c). In the above type of recoil experiment,
if performed above the $\pi^+\pi^-$ threshold, some background will be due
also to the non-interfering processes of fig.1d).
Magnitudes of the signal and the background are discussed below. These are also
used to justify the coalescence model and to find its limitations.

At first the two-meson subsystem in fig.1 is discussed. Then later this
is extended to the three body situation. Details of the calculations as
well as some definitions are collected in Appendix 1. As will be shown below,
the atom is formed predominantly via the charged intermediate state of
fig 1a) i.e. as an intermediate state in the reaction
$\pi^+\pi^-\rightarrow \pi^0\pi^0$. The relevant amplitude is the product
of three factors $F_c \cdot G^c_c \cdot \hat{T}_{co}$, where $F_c$ is an
amplitude for the
 production of charged mesons and $G^c_c$ is a coulomb propagator for the
closed channel (c) that
-- close to the 1s-atomic state -- is dominated by
$|\psi><\psi|/(E-\epsilon_0)$. The last factor ($\hat{T}_{co}$) is a transition
matrix, which for reasons of normalisation, is put into a pseudopotential
form $\hat{T}=\frac{2\pi}{\mu}\delta({\bf r})T$. The detailed properties and
the
parametrisation of $T$ are given in Appendix 1, where it is shown that close
 to the atomic state pole $T_{co}$ is
\be
\label{4.1}
T_{co}\approx \frac{A_{co}}{1+A_{cc}R/(E-\epsilon_0) },
\ee
where $R=-\frac{2\pi}{\mu}|\psi(0)|^2$. Collecting together the three factors
results in
the atomic production amplitude
\be
\label{4.2}
F_{co}=F_c\cdot G^c_c\cdot \hat{T}_{co}=
\int d{\bf r}F_c(r)\psi({\bf r})\cdot \frac{2\pi}{\mu}\cdot \frac{A_{co}}
{E-\epsilon_0+RA_{cc}}\int
d{\bf r'}\psi({\bf r'})\delta({\bf r'}),
\ee
where $A_{cc}$ is the scattering length in the $\pi^+\pi^-$ channel and
$A_{co}$ is the transition length responsible for the atomic decay into
the $\pi^0\pi^0$ channel. The distances involved in $F(r)$ are much shorter
than the atomic radii and so the first integral reduces to
$\psi(0)\cdot F_c$, where $F_c=\int d{\bf r}F_c(r)$ is now the $\pi^+\pi^-$
production amplitude. The atomic formation amplitude becomes
\be
\label{4.3}
F_{co}=- F_c \cdot \left(\frac{RA_{co}}{E-\epsilon_0+RA_{cc}}\right)
\ee
and displays the standard Breit-Wigner shape for the line. This does not happen
 in the case of the other production amplitude due to the
$\pi^0\pi^0\rightarrow \pi^0\pi^0$ reaction of fig. 1b). By an analogous
calculation one obtains
\be
\label{4.4}
F_{oo}=F_o\cdot G_o\cdot \hat{T}_{oo}=-\int d{\bf r}F_o(r)\frac{e^{i{\bf
q_o.r}}}{r}
\left[\frac{K_{oo}}{1+iq_oK_{oo}}
\left(\frac{E-\epsilon_0+R(K_{cc}-\frac{K^2_{oc}}{K_{oo}})}
{E-\epsilon_0+RA_{cc}}\right)\right],
\ee
where now $F_o(r)$ is an amplitude for $\pi^0\pi^0$ production to be integrated
over the intermediate state propagator. The terms in brackets come from
$T_{oo}$ expressed by eq.(\ref{A2}) and expanded around the atomic pole
 with the help of eq.(\ref{A6}) and some simple but lengthy algebra.
The terms in the round brackets constitute another atomic signal that
now displays a typical Fano resonance consisting of a peak and a {\bf zero}
in the amplitude \cite{Fano}. The position of the peak is given by the
atomic level energy and the separation between the zero and the peak is a
measure of  the $K_{oc}$ coupling strength relative to the $K_{oo}$ and
$K_{cc}$ strengths. The pionium signals in these two channels are plotted
in fig.2a). They are  clearly too narrow to be measured directly at the present
time. A similar structure is also predicted for the $(K^+\pi^-)$-atom.
However, in the $K\bar{K}$ system -- see table 1 and fig.2b) -- the peaks are
expected to have  widths of the order of keV and may be even broader in higher
$Z$
systems. Such reactions with an initial proton beam of 3keV resolution
could possibly permit a direct measurement of this shape\cite{Nann}.

The contribution of the elastic $\pi^0\pi^0$ scattering to pionium
production  from eq.(\ref{4.4}) is small.
One reason is the $<K_{oo}/r>$ factor, which is
$\approx 0.1$. Another is that the $\pi^+\pi^-$ production cross-section is
larger by a factor of about 3--4. The lower estimate comes from two pion
production cross-sections at $p_{lab}$=2.23GeV/c \cite{Hera}.  In fact
as $p_{lab}$ decreases this factor increases. The upper
estimate comes from the coupling ratio $G(pn\pi^+)/G(pp\pi^0)=\sqrt{2}$
\cite{Dumb}.
This is a fortunate conclusion, since the $F_{oo}$ amplitude depends on the
off-shell structure in $F_o(r)$. However, from fig.1c) the direct production
of $\pi^0\pi^0$ described by simply $F_o$ contributes
both background and interference terms to the atomic amplitude $F_{co}$. To
study these effects one needs 3-body kinematics appropriate for the final
states in fig. 1.  Let the energy be measured relative to the the
$\pi^+\pi^-$ threshold, which for the final state in the cm system is
\be
\label{4.5}
E_c=q_o^2/2\mu-\Delta+E_R(P),
\ee
where $E_R(P)=P^2/2\mu_R$ is the $^3$He recoil energy and
$\Delta=2(m_+-m_0)$ is the threshold mass difference. Neglecting the $F_{oo}$
contribution from eq.(\ref{4.4}), since it is expected to be only a $5\%$
correction to $F_{co}$, the formation probabilities  are given by
the integration
\be
\label{4.6}
W=\int |F_{co}+F_o|^2dL^3=\int \{|F_o|^2 +[|F_{co}+F_o|^2-|F_o|^2]\}dL^3
=W_B+W_A
\ee
over the three body final phase space. This is done in Appendix 2. The second
term on the RHS gives the probability of atom formation $W_A$, whereas
the first term is the $ \pi^0\pi^0$ background.

Integration over the atomic line and the phase space yields the total atom
formation probability
\be
\label{4.7}
W_A=\frac{|\psi(0)|^2}{2\mu} \cdot |F_c|^2\cdot L^2 \cdot
\left(1+2q_o\frac{ {\rm Im} (\lambda A_{co})}{|\lambda|^2}\right),
\ee
where $L^2$ is the phase space element $L^2=P_A/(4\pi(2m+M_3))$ with

\noindent$P\rightarrow P_A=\sqrt{2\mu_R(|\epsilon|+E_c)}$
 being the recoil momentum at the atom production energy and $M_3$ is the mass
of $^3$He. The last factor in eq.(\ref{4.7}), where $\lambda=F_o/F_c$, is
due to the atomic background interference. This term is an improvement over
 the coalescence model estimate given by eq.(\ref{3.3}), but the correction
is negligible $(<10^{-3})$ in the pionium case. Similarly in the $K^-\pi^+$
 case Im$ A_{co} \approx 10^{-4}$ -- again a negligible correction.
However, it may be of the
order 1 in the $K\bar{K}$ case. In general, the experimental resolution
will be such that many atomic states will fall inside this resolution. In
this case the above factor $|\psi(0)|^2$ is replaced by
$\sum_n |\psi_n(0)|^2$, where $n$ is the principal quantum number.
This can lead to an enhancement that is, at most, $20\%$.

The recoil energy distribution due to direct $\pi^0\pi^0$ production requires
 one integration less and is given by
\be
\label{4.8}
W_B(E_R)=|F_o|^2P(E_R)\sqrt{2\mu((E_c+\Delta)-E_R)}/[16\pi^3(2m+M_3)].
\ee
A similar expression holds for the recoil due to
$\pi^+\pi^-$ production, if experiments are performed above that threshold.
The necessary change is $F_o\rightarrow F_c \ \ , \ \ E_c\rightarrow
E_c-\Delta$.
The latter effect of the $\pi^+\pi^-$ channel may be minimised by making
the experiment close to the threshold \cite{Nann}.  However, in the energy
range of interest, the $\pi^0\pi^0$
production makes a constant but large background. With the energy
 fixed at the $\pi^+\pi^-$ threshold one obtains
$W_A/[W_B(E_R)\Delta E_R]\approx |F_c/F_o|^2/200$, where a recoil energy
resolution of 70keV was used\cite{Nann}.
As discussed above, the enhancement due to $|F_c/F_o|^2$ is expected to be at
least a factor of 3.  In spite of this, one should expect a sizeable background
in the
subthreshold pionium production. Only if the resolution can be reduced
substantially from the stated value of 70keV will the atomic state signal
become significant.

Kaonium production is likely to be dominated by the $f_0(975)$ and
$a_0(983)$ resonances. Indeed, a peak is observed at Saturne in the reaction
p(d,$^3$He)X around the $K\bar{K}$ threshold\cite{Siebert}. The
$K^+K^-$ atom production rate and the background recoil energy distribution
 due to $\pi\pi $ or $\eta\pi$ are given by equations analogous to
eq.(\ref{4.7},\ref{4.8}), but with relativistic phase space. Now one
obtains

\noindent $W_A/[W_B(E_R)\Delta E_R]\approx |F_c/F_o|^2/225$, where
$F_c/F_o$ is the ratio of resonant decays into $K\bar{K}$ versus
$\pi\pi$ or $\eta\pi$. This factor is available in the $f_0(975),I=0$ case
and $|F_c/F_o|^2=g_K/g_{\pi}=2$ -- see ref.\cite{fdecay}. Unfortunately,
there is no experimental estimate of the $a_0(983),I=1$ coupling to
$K\bar{K}$ relative to the $\eta\pi$ coupling. However, in some theoretical
models,
a ratio similar to that in the $f_0$ case is suggested\cite{Isgur}.
In view of the strong $K\bar{K},\pi\pi$ coupling to the resonant states, the
resultant scattering lengths and K-matrix elements could be as large as 1--2
fm.
This generates structure in the atomic line with a width of the order keV.
A full set of K-matrix elements is not yet available and so a more
detailed structure cannot be calculated. In particular, one cannot estimate
with reliable parameters
the magnitude of the Fano type of structure already predicted for pionium in
fig.2a). For illustration in fig. 2b), reasonable values for the K-matrix
suggested in ref.\cite{Can} are used. However, it should be emphasised that
the detailed structure seen in this figure is very dependent on the actual
numbers used in the K-matrix .
The interference effect in eq.(\ref{4.7}) may be rather large, since
$|2q_0A_{co}|\approx 4$ is expected.

The $K^+\pi^-$ case is rather similar to that of pionium in the sense that the
threshold of the open channel $K^0\pi^0$ is close to the energy of the atom.
However, the major difference is that a suitable mechanism for producing
$K^+\pi^-$ or $K^0\pi^0$ is not known -- unlike:

a) Pionium, which is part of the well known two-pion production present in
many reactions e.g. as in eq.(\ref{1.1}), or

b) Kaonium, which can be produced via the $f_0(975)$ and $a_o(983)$ resonances.

\noindent One possibility is to produce $K\pi$ in the reaction
$pp \rightarrow p\Sigma^+ \pi^-K^+$ -- but this has a very small cross-section
e.g. 0.019mb at 3.47GeV/c \cite{Hera}.

\vskip 1.0 cm

\section{ The production of exotic atoms in 2p-states}
\setcounter{equation}{0}

The formation of pionium in a 2p-state would make its detection much easier.
The main decay mode is the X-ray transition to the 1s-state with a rate
of $0.86\cdot 10^{11} s^{-1}$ \cite{Dumb1}. This time is long enough for the
atom to leave the target and so facilitate measurements of its decay modes.
However, its production rate is {\em very} slow, since it involves an
additional factor of $\alpha^2$. Consider that a vector meson, e.g.
$\rho$ or $\phi$, produced by a reaction such as
$pd\rightarrow \rho \ ^3He$ decays into pionium or kaonium. Let the
meson production amplitude be $F{\bf \epsilon}\cdot {\bf q}$, where
${\bf \epsilon}$ is a unit vector related to the vector particle and
${\bf q}$ is the momentum in the meson-meson centre-of-mass.
Employing a procedure
analogous to that used in section 2 yields an atom formation rate
 in a coalescence model  of the form
\be
\label{5.1}
\frac{W_A}{W_B}=\frac{3}{4\pi}|\psi'_r(0)|^2\frac{1}{2\mu}
\frac{L^{n+1}}{<q^2L^{n+2}>}.
\ee
Here, $\psi'$ is the radial derivative of the p-wave atomic function at the
origin -- $\sqrt{(\mu \alpha)^5/24}$ for
a 2p-state -- and $<...>$ denotes the average
of $q^2$ over the final phase space. Close to the $\pi^+\pi^-$ production
threshold this ratio follows $E_c^{-5/2}$ and, in principle, may be quite
large. However, in practice, vector meson production may be minute.
Estimates suggest that, at $E_c$=10MeV, the ratio  $W_A/W_B(E_R)\Delta E_R$
 is of the order $10^{-9}$.
Similar calculations may be carried out for estimating the chances to produce
kaonium in the reaction $e^+e^- \rightarrow \phi \rightarrow (K^+K^-)_{2p}$.
Right at the $(K^+K^-)$ atomic energy, one has
$W_A/[W_B(E)\Delta E]\approx 0.4\cdot 10^{-6}$ using the DAFNE resolution
of $4\cdot 10^{-4}$. This reaction is 32MeV away from the $\phi$ energy, which
gives another 1/500 reduction factor. With the expected rate of
$10^4 \phi$/sec, this would produce about one kaonium/day.

\vskip 1.0 cm

\section{ Conclusion}
\setcounter{equation}{0}

In this article, signals indicating the presence of the exotic atoms
pionium ($\pi^+\pi^-$), kaonium ($K^+K^-$) and $K^+\pi^-$ are estimated in
various reactions.
The emphasis has been on the production of pionium in the reaction \\
$ p+d \rightarrow \  ^3He+(M^+M^-)_{Atom}$, but the formalism proposed applies
equally well to kaonium and the $K^+\pi^-$ atom. Furthermore, following the
suggestion in ref.\cite{Nann}, the signal in the $^3$He recoil energy
distribution is expressed as the ratio of probabilities
$W_A(Atom)/W_B(Background)\Delta E_R$, where $\Delta E_R$ is
the recoil energy resolution. In the pionium case, since this resolution
is 70keV compared with the atomic line width of a few eV, the signal is
only $\approx 1\%$ of the $\pi^0\pi^0$ background. Only if the resolution
can be decreased
by an order of magnitude could such an atomic signal become measureable.

The situation for seeing the $K^+\pi^-$ atom is somewhat similar to that of
pionium
 with the additional complication that the basic $K\pi$ states
are more difficult to generate i.e. in eqs.(\ref{4.3},\ref{4.4}), the
$F_{c,o}(K\pi)$ factors are much smaller than the corresponding
$F_{c,o}(\pi\pi)$ factors.

Possibly it is kaonium that will be most easy to observe, since this has a
width that is of the order 1 keV compared with fractions of an eV for pionium
and the $K^+\pi^-$ atom -- see table A1. In addition, the basic mechanism for
creating this atom  is enhanced, since it appears via the $f_0(975)$ and
$a_0(983)$ resonances. In comparison, the production of, for example,
the $\rho$-meson would generate pionium only in a {\em 2p-wave} and this would
be expected to have a very small probability. It is possible that the peaks
seen at the $K\bar{K}$ threshold in the $p(d, \ ^3He)X$ reaction at
Saturne are already partly due to kaonium formation. Unfortunately, unlike
the pionium and $K^+\pi^-$ systems, the theoretical estimates based on
eqs.(\ref{4.7},\ref{4.8}) can not
yet be made, since they require a knowledge of the K-matrix elements
$K_{cc}(K^+K^- \rightarrow K^+K^-), K_{oo}(\pi\pi \rightarrow \pi\pi)$ and
$K_{co}(K^+K^- \rightarrow \pi\pi)$. For pionium and $K^+\pi^-$, the
corresponding K-matrix
elements are sufficiently well known to show that their effect -- reflected
by the interference term in eq.(\ref{4.7}) -- represents only a $10^{-3}$
correction
to the basic coalescence model. On the other hand, for the $K\bar{K}$
system, all that can be said at present is that Im$A_{cc} \approx -1$fm, which
is sufficient to indicate that the interference term in eq.(\ref{4.7}) could
be significant.
Similarly, the effect of the $K^+K^-$ atom in the background production
of fig.1b) [i.e. eq.(\ref{4.4})] could also be significant, if Re$A_{cc}$
 and also
the K-matrix elements are sufficiently large. However, it is possible that
these parameters could conspire to give a Fano zero that greatly reduces
the overall effect.

As outlined in section 5, the production of atoms in the 2p-state, although
desirable for observing such atoms when once generated, is very unlikely.
Similarly, the direct production via
$e^+e^- \rightarrow \phi \rightarrow (K^+K^-)_{2p}$ seems also to be
impractical.

So far the discussion has concentrated on  $p+d$ reactions with the conclusion
that the signals for atom production will be very difficult to detect compared
with the ever present background. In $pp\rightarrow pp\pi^+\pi^-$ reactions
this
problem is further complicated by the need to consider 3 and 4 phase space
factors i.e. n=2 in eq.(\ref{3.3}).
The relative atomic production rates W(atom)/W(pair) in the
   $pp\rightarrow pp \pi^+ \pi^-$ and $ K \bar{K}$
 or in the $pp\rightarrow  K^- \pi^+ p \Sigma^+$
    reactions are now higher . Eq.(\ref{3.3}) produces a rate of
$1.26\times 10^{-5}[2 \mu_{M^+M^-}/(E-E(threshold))^{3/2}]$ that is more than
    $10^{-2}$ at 1 MeV or $10^{-4} - 10^{-3}$ at 10 MeV above the thresholds .
However, it would then be necessary to measure
the energies $(E_1,E_2)$ of both protons and to concentrate on the
events with $E_1+E_2$ tuned to pick up the atomic signal. The ratio of the
atomic formation probability to that of uncorrelated mesons would now have
the form $W_A/W_B(E_{p_1},E_{p_2})\Delta E_{p_1} \Delta E_{p_2}$ i.e. with
the resolution appearing twice. This is clearly a much more demanding
experiment than the $p+d$ proposal.

One of the authors (S.W.) wishes to acknowledge the hospitality of the
Research Institute for Theoretical Physics, Helsinki, where part of this
work was carried out. In addition, he wishes to acknowledge the receipt
of the KBN grant number 2p 302 14004, which partially covered the expenses
incurred
in this collaboration.  Also the authors wish to thank Drs. B.H\"{o}istad ,
H.Nann, W.Oelert and J.Stepaniak for useful communications and discussions.

\vskip 1.0 cm

\appendix

\chapter{Scattering Parameters}
%\setcounter{equation}{0}

%\section*{Scattering Parameters}
\renewcommand{\theequation}{A.\arabic{equation}}

\vskip 1.0cm
As there exists conflicting sign conventions for the low energy expansions
a few basic definitions seems appropriate. The scattering matrix is
normalised according to the Lippman-Schwinger equation
$\hat{T}=\hat{V}+\hat{V}(E_c-\hat{H}_0)^{-1}\hat{T}$ that leads to a low energy
parametrization $1/T=1/a+iq$, where $a$ is a scattering length and $q$ is
a relative momentum. Phenomenological scattering lengths are used and an
off-shell extension of this model is obtained by a pseudopotential
$\hat{T}=\frac{2\pi}{\mu} \delta({\bf r})T$ that may generate a multiple
scattering
expansion. Scattering lengths defined in this way $a=-tan\delta/q$ differ
in sign from those defined in the compilation of ref.\cite{Dumb}.

Consider s-wave scattering in a system consisting of two channels $c$ and $o$.
The $K$-matrix, i.e. a generalisation of the scattering length, and the
$T$-matrix that follows from it are denoted by

\be
\label{A1}
  \hat{K} = \left( \begin{array}{ll}
 K_{oo} & K_{co} \\
 K_{oc} & K_{cc} \end{array} \right)
 \ \ \ {\rm and} \ \
  T = \left( \begin{array}{ll}
 \frac{A_{oo}}{1+iq_oA_{oo}}&  \frac{A_{co}}{1+iq_cA_{cc}}\\
  \frac{A_{oc}}{1+iq_cA_{cc}}&  \frac{A_{cc}}{1+iq_cA_{cc}}\end{array} \right),
\ee
where $q_{o,c}$ are the centre-of-mass momenta of the two mesons in the two
channels $o,c$.
The channel scattering lengths $A_{ij}$ are expressed in terms of the
$K$-matrix elements, via the solution of $T=K-iKqT$,  by \cite{Pilk}

\[ A_{cc}=K_{cc}-iK^2_{co}q_o/(1+iq_oK_{oo}) \]
\[ A_{co}=K_{co}/(1+iq_oK_{oo}) \]
\be
\label{A2}
A_{oo}=K_{oo}-iK^2_{co}q_c/(1+iq_cK_{cc})
\ee
These equations form a basis in which to describe two channel scattering
in terms of the three parameters of the $K$-matrix.

The systems of interest, like $\pi\pi$ or $K\pi$, are believed to display
a high degree of isospin symmetry in their interactions. The $K$-matrix
is given by two real parameters referring to isospin 0, 2 in the $\pi^+\pi^-$
case and 1/2, 3/2 in the $K^+\pi^-$ case. The isospin structure of the
$K$-matrix is given in table A1.

The case of $K^+K^-$ is more complicated due to the mass difference of
the $K^+K^-$ and $\bar{K}^0K^0$ and also the presence of the two decay channels
 $\pi\pi$(I=0) and $\eta\pi$(I=1).

Also notice the unitarity condition, which follows from eqs.(\ref{A2}),
\be
\label{A3}
Im A_{cc}=-|A_{co}|^2q_o.
\ee

Isospin symmetry is broken by coulomb interactions and meson mass differences.
The first allow for atomic binding and the second induce decays of these
atomic systems. The channels in table A1 are, accordingly, named closed (c)
and open (o).

To describe atomic systems produced in collisions, these effects have to be
built into eqs.(\ref{A1},\ref{A2}). The standard way is to separate long
range effects that enter coulomb propagators $G^c$ and modifications at short
ranges entering "coulomb corrected" scattering lengths. The propagators that
describe coulomb and short ranged interactions are given by
$G^c+G^cTG^c$. As a consequence, the scattering is described by amplitudes
of the form
\be
\label{A4}
f_{ij}=f^c\delta_{ij}+e^{i\sigma_i}c^2_iT_{ij}^c c^2_je^{i\sigma_j},
\ee
where $f^c, \ \sigma$ and $c$ are the coulomb scattering amplitudes, phases
and penetration factors. The latter arise in asymptotic solutions only and
are of no concern in the present problem. The coulomb corrections to $T^c$
arise in terms of known coulomb functions that should replace $iq$ in
eqs.(\ref{A1},\ref{A2}):
\be
\label{A5}
iq\rightarrow f=2\gamma h+iqc^2,
\ee
\[ {\rm where} \ \ \ \ c^2=2\pi\eta/[\exp(2\pi\eta)-1]  \ \ \ {\rm and}
 \ \ \ \ \ h=\frac{1}{2}[\psi(i\eta)+\psi(-i\eta)]-\frac{1}{2} {\rm
ln}\eta^2.\]
Also $\gamma=ZZ'\alpha\mu$ and $\eta=\gamma/q$ -- see for example
ref.\cite{New}.
 All these functions are
diagonal in channel indices, which were dropped however. It is easy to check
that for neutral particles ($Z=Z'=0$) the $f=iq$. Now, the atomic states
created in the intermediate states of nuclear reactions are generated by
poles in $f$ of eq.(\ref{A5}). The expansion of $f$ around  this pole
yields \cite{Carbon}
\be
\label{A6}
f\simeq\frac{R}{E_c-\epsilon_0}, \ \ {\rm where} \ \
R=-\frac{2\pi}{\mu} |\psi(0)|^2 = -2\mu^2\alpha^3,
\ee
$\epsilon_0$ is the pure coulomb energy of an s-level and $E_c$ is the energy
distance from the $M^+M^-$ threshold. As a first application of eq.(\ref{A5}),
the position of the atomic pole in $T_{cc}$ of eq.({\ref{A1}) is found. One
obtains immediately the atomic energy shifted by strong interactions

\be
\label{A7}
\epsilon-i\Gamma/2=\epsilon_0+\frac{2\pi}{\mu}|\psi(0)|^2A_{cc}
=\epsilon_0+2\mu^2\alpha^3A_{cc},
\ee
which is a well known result relating level shifts and widths to the
scattering length. The width expressed by $K$-matrix elements is
\be
\label{A8}
\Gamma/2=2\mu^2\alpha^3\frac{K^2_{oc}q_o}{(1+q_o^2K^2_{oo})}
\ee

\vskip 1.0 cm

%\section*{Phase space factors and integrals}
\chapter{Phase space factors and integrals}
\renewcommand{\theequation}{B.\arabic{equation}}
\setcounter{equation}{0}

In this appendix are calculated phase space factors and integrals over
the atomic line. Non-relativistic energies are used for the
$^3He \  \pi^0\pi^0$ final state. For a final state containing three
particles
\be
\label{B1}
dL^3=(2\pi)^4\delta(\sum_i {\bf p_i})\delta (E_0-\sum_i p_i^2/2m_i)
\prod_i\frac{d{\bf p_i}}{(2\pi)^32m_i}.
\ee
Going over to energy variables $E_q=q^2/2\mu$ in the $\pi\pi$ centre-of-mass
system
and the recoil energy $E_R(P)=P^2/2\mu_R$ with $\mu_R=2mM_3/(2m+M_3)$,
eq.(\ref{B1}) becomes
\be
\label{B2}
dL^3=\delta(E_0-E_q-E_R)PqdE_RdE_q/c,
\ee
where $c=16\pi^3(2m+M_3)$. Thus the background $\pi^0\pi^0$ emission generates
the recoil energy distribution given by
\be
\label{B3}
W_B(E_R)=|F_o|^2\int \delta(E_0-E_q-E_R)PqdE_q/c=
|F_o|^2P\sqrt{2\mu(E_0-E_R)}/c.
\ee
Here $E_0$ is the energy related to the $\pi^0\pi^0$ threshold. Now the
atomic formation probability is calculated  and
$E_c=E_0-2(m_+-m_0)=E_0-\Delta$ i.e. the energy related to the $\pi^+\pi^-$
threshold is more appropriate for use.

The atomic signal from eq.(\ref{4.6}) using eq.(\ref{4.3}) becomes
\[W_A=|F_o|^2\int \left[ |\frac{\lambda RA_{co}}{E_q-\Delta-\epsilon_0+RA_{cc}}
-1|^2-1\right] \delta(E_0-E_q-E_R)P q dE_RdE_q/c\]
\be
\label{B4}
=|F_o|^2\int \left[ |\frac{\lambda RA_{co}}{E_0-E_R-\Delta-\epsilon_0+RA_{cc}}
-1|^2-1\right] P q dE_R/c,
\ee
where $\lambda=F_c/F_o$ and $q=\sqrt{2\mu(E_0-E_R)}$. Since the atomic line
is so narrow, it is assumed that $q$ is constant over the energy range of
interest.
 The integration over the atomic line is now performed using the relation
    $\int dE_R/[(E_R-\epsilon)^2+(\Gamma/2)^2]=2\pi/\Gamma$ and also by
taking into account the unitarity condition
of eq.(\ref{A3}) and replacing $R$ by eq.(\ref{A6}). This results in

\be
\label{B6}
W_A=|F_c|^2\frac{|\psi(0)|^2}{2\mu}\left(1+\frac{2q_o({\rm Im}\lambda A_{co})}
{|\lambda|^2}\right)\left[\frac{P_A}{4\pi(2m+M_3)}\right],
\ee
where the first term comes from the line and the second represents
interference with the background.
The square bracket term is the $L^2$ phase space factor for the atom
and the $^3$He system, $P_A=\sqrt{2\mu_R(E_c-\epsilon)}$ is the recoil
momentum at the atom production point and $q_o$ is the relative momentum of
the $\pi^0\pi^0$ emitted in the atomic decays.

\newpage

\vskip 0.5cm

Table 1 Some properties of the exotic atoms
$\pi^+\pi^-$ , $K^+\pi^-$ and $K^+K^-$.

\vskip 0.5cm

\begin{tabular}{|c|c|c|c|}                          \hline
           & $\pi^+\pi^-$ & $K^+\pi^-$ & $K^+K^-$ \\      \hline
Bohr radius/fm& 387.46     & 248.51 & 109.55     \\ \hline
$E_{1s}$/keV  & --1.86   & --2.90 &  --6.57     \\ \hline
$E_{2p}$/keV  & --0.46  &  --0.72  & --1.64           \\ \hline
Basic Decay from 1s state &$\pi^0\pi^0$& $K^0\pi^0$ & $\pi\pi$ and $\eta\pi$\\
\hline
Width (ev)&0.24&0.18&639.4 \\
Lifetime ($10^{-15}$sec)&2.72&3.59&0.0010 \\ \hline \hline
\end{tabular}

\vskip 0.5cm
Table A1 The isospin structure of the K-matrix

\vskip 0.5cm

\begin{tabular}{|c|c|c|c|}                          \hline
Channel \  c& $\pi^+\pi^-$& $K^+\pi^-$& $K^+K^-$ \\
\ \ \ \ \ \ \ \ \ \ \ \ \ o&$ \pi^0\pi^0$& $K^0\pi^0$ & $\pi\pi$, $\eta\pi$ \\
\hline
$K_{cc}$    & $(a_2+2a_0)/3$ & $(a_3+2a_1)/3$ & $(A_0+A_1)/2$ \\
$K_{co}$ & $\sqrt{2}(a_2-a_0)/3$ & $\sqrt{2}(a_3-a_1)/3$ & \\
$K_{oo}$ & $(2a_2+a_0)/3$ & $(2a_3+a_1)/3$ & \\ \hline
\end{tabular}

\vskip 0.5cm

\newpage
{\bf \ \ \ \  Figure Captions}

\vskip 0.5cm

Figure 1. The basic amplitudes for atom formation and background above and
below the $\pi^+\pi^-$ threshold:

\vskip 0.5cm

a) Direct atom   \ \ \ \ \ \ \ \ \ \ \ \ \ \ b) Fano resonance.

c) Direct $\pi^0\pi^0$ background \ \ \ d) Charged background

\vskip 0.5cm

Fig.2 Shape of the 1s-atomic state contribution to the background
amplitude. The moduli squared of the round bracket terms (...) in
eqs.(\ref{4.3},\ref{4.4}) are plotted -- the dashed line corresponds
to eq.(\ref{4.3}) and the solid line to eq.(\ref{4.4}).

a) The effect of pionium. The $\pi^0\pi^0$ contribution
of eq.(\ref{4.4}) exhibits a Fano resonance with its characteristic zero.
Here the values of $a_0$ and $a_2$ quoted in section 2 are used.
N.B. The semi-log scale .

b) The effect of kaonium. For illustration, reasonable values
of $K_{oo}=0.328$, $K_{cc}=2.53$ and  $K_{co}=0.948$ all in fm. are
used -- see  ref.\cite{Can}. For clarity, the dashed line has been
multiplied by a factor of 10.
\vskip 0.5cm

\end{document}